\documentclass{elsart}

\usepackage{graphics}
\usepackage{amsmath}
\usepackage[dvips]{graphicx}
\usepackage{amsfonts,amsbsy}
\usepackage{amssymb}
\usepackage{bbm}
\newcommand{\beq}{\begin{equation}}
\newcommand{\eeq}{\end{equation}}
\newcommand{\beqa}{\begin{eqnarray}}
\newcommand{\eeqa}{\end{eqnarray}}
\def\r{{\boldsymbol r}}
\def\z{{\boldsymbol z}}
\def\x{{\boldsymbol x}}
\def\y{{\boldsymbol y}}
\def\k{{\boldsymbol k}}
\def\q{{\boldsymbol q}}
\def\p{{\boldsymbol p}}
\def\0{{\boldsymbol 0}}
\def\v{{\boldsymbol v}}
\def\nab{{\boldsymbol \nabla}}

\def\cal{\mathcal}

\def\bra#1{\langle#1\vert}
\def\ket#1{\vert#1\rangle}

\begin{document}
\begin{frontmatter}

\title{Real and imaginary-time $Q\overline{Q}$ correlators in a thermal medium}
\author{A. Beraudo$^1$, J.P. Blaizot$^1$ and C. Ratti$^{1,2}$}
\address{$^1$ECT*,
  strada delle Tabarelle 286, I-38050 Villazzano (Trento), Italy;\\
$^2$Department of Physics and Astronomy, SUNY Stony-Brook, NY 11794.}

\date{\today}

\begin{abstract} 
We investigate the behavior of a pair of heavy fermions, denoted by $Q$ and $\bar{Q}$,    in a hot/dense medium. Although we have in mind the situation where $Q$ and $\bar{Q}$ denote heavy quarks, our treatment will  be limited to simplified models, which bear only some general similarities with QCD. We study in particular the limiting case where the mass of the heavy fermions is infinite. Then a number of results can be derived exactly: a Schr\"odinger equation can be established for the correlator of the heavy quarks;  the interaction effects exponentiate, leading to a simple instantaneous effective potential for this Schr\"odinger equation. We consider simple models for the medium in which the $Q\bar Q$ pair propagates. In the case where the medium is a plasma of photons and light charged fermions, an imaginary part develops in this effective potential. We discuss the physical interpretation of this imaginary part in terms of the collisions between the heavy particles and the light fermions of the medium; the same collisions also determine the damping rate of the heavy fermions. Finally we study the connection between the real-time propagator of the heavy fermion pair and its  Euclidean counterpart, and show that the real part of the potential entering the Schr\"odinger equation for the real-time propagator is the free energy calculated in the imaginary-time formalism. \end{abstract}
\end{frontmatter}
%
\section{Introduction}%

The propagation of heavy quark bound states, such as the $J/\Psi$,  in a hot and dense medium has been the subject of many studies, since the original  proposal by Matsui and Satz \cite{satz} 
suggesting that the suppressed production of such bound states  in ultra-relativistic heavy-ion collisions could be used as a signal of
deconfinement. The initial suggestion \cite{satz} assumed that the dominant effect of the interaction with the medium is the screening of  the  $Q\bar{Q}$ interaction potential, leading to a disappearance  of the bound states when the screening is sufficiently strong. The screened potential can be obtained from equilibrium calculations of the force between two infinitely heavy color charges, a quantity which can, in principle, be extracted from the lattice data for the
correlator of two Polyakov lines. 
Such a correlator is in fact  related to
the change of free energy occurring when a heavy
$Q\overline{Q}$ pair is added into a hot QCD medium \cite{mclerr}, and the relation between this  free-energy and the effective 
potential to be  inserted in a Scr\"odinger equation
(or in a T-matrix calculation, like in \cite{rapp1,rapp2})
remains an open question. Thus, the effective potential has been identified alternatively with the $Q\overline{Q}$ free
energy \cite{diga1,diga2,wong1,wong2}, the internal energy \cite{shury,ber1}
or even with a linear combination of both  \cite{wong3,wong4,ber2}, leading 
to different dissociation temperatures for the $J/\Psi$,
depending on the choice made.
 
In another line of investigation one considers the
spectral functions of a $c\bar{c}$ and $b\bar{b}$ mesons in a hot  environment \cite{dat,asa1,asa2,aarts,petre1}.
Such objects
contain in principle the full physical information on the properties of the mesonic 
excitations in a medium, 
and are also directly related to observable quantities, like the dilepton
production rate in the vector channel.
Unfortunately spectral functions are not directly measured in a lattice simulation,
but have to be reconstructed
from Euclidean 
correlators, and this reconstruction is plagued with large
uncertainties. To go around such difficulties, one may attempt to compare directly the spectral functions and the corresponding imaginary-time correlators that are obtained from lattice calculations to those that can be calculated from a Schr\"odinger equation with a screened potential. Such studies were carried out recently \cite{rapp2,ber2,petre2,mocsy1,mocsy2,ber3}.

More recently, still  another approach was developed, in which the focus is on the \emph{real-time} propagator of a heavy-quark pair. In Refs.~ \cite{lai1,lai2,lai3,lai4,lai5}, it was argued that such a propagator obeys a 
 Schr\"odinger  equation 
with an effective potential which was derived first  perturbatively, and then estimated in classical lattice gauge simulations. An interesting observation made in \cite{lai1} is that this effective potential may contain an imaginary part that may strongly affect the calculated spectral function. 

In this paper we shall discuss a number of issues triggered by the latter studies. We shall examine under which conditions, the
in-medium $Q\overline{Q}$-propagator 
obeys a Schr\"odinger equation. To do so, we shall consider the general equation of motion satisfied by the heavy fermion pair correlator. This is part of an infinite hierarchy of coupled equations for the $n$-point functions. However,  in the limit where the mass of the heavy fermion is infinite, the equation for the $Q\overline{Q}$ correlator decouples and reduces to a closed, Schr\"odinger equation.  We shall also establish a link beween the effective potential entering this Schr\"odinger equation
and the heavy-quark free-energy that can be extracted from lattice data; in other words, we shall investigate the relations between the real-time and imaginary-time correlators, and understand the origin of the imaginary part  found in \cite{lai1,lai2,lai3,lai4,lai5}.

The discussion will be general and pedagogical: our purpose is to illustrate simple and general physical effects, rather than to obtain quantitative results. We shall focus on a specific correlator, which is defined in   Sec. \ref{sec:basic}, where we also recall basic properties of real and imaginary time correlators and their relation through analytic continuation. Then we discuss several models for the medium in which the heavy quarks propagate. The medium is assumed to be a non relativistic cold Fermi gas in Sec. \ref{sec:nrtoy}, a thermal bath of mesons in Sec.~ \ref{sec:scalartoy} and a thermal bath of photons and charged particles in  Sec.~\ref{sec:QEDtoy}. The interaction between the two heavy fermions is an instantaneous potential in the first case, is mediated by meson exchange in the second case, and photon exchange in the latter. We shall see that in the limit where the mass of the heavy fermions becomes infinite, the correlator obeys indeed, for large times, a Schr\"odinger  equation with an effective potential that accounts for infinite resummation of static interactions. 
In Sec.~\ref{sec:QEDtoy}, we show that the effective potential can develop an imaginary part corresponding
to scattering processes between the heavy quarks and the light fermions
of the thermal bath. 
Finally, in Sec. \ref{sec:concl} we summarize our results.
\section{Some definitions}
\label{sec:basic}

We shall consider a pair of heavy fermions, denoted respectively by $Q$ and $\bar Q$, which are supposed to be sufficiently heavy that they can be described by non relativistic quantum mechanics.  We shall denote the quantum fields associated to $Q$ and $\bar Q$ respectively by $\psi$ and $\chi$. We shall often refer to $Q$ and $\bar Q$ as to a quark and an antiquark, although,  in the simplified models that we shall discuss, these will not necessarily have all the attributes of the quarks of QCD. 

We shall focus our discussion  on the evaluation of the following
 propagator:
\beq
G^>(t,\r_1;t,\r_2|0,\r_1';0,\r_2')\equiv
\langle \chi(t,\r_2)\psi(t,\r_1)
\psi^\dagger(0,\r_1')\chi^\dagger(0,\r_2')\rangle,\label{eq:2parta}
\eeq
which represents the probability amplitude to find, at time $t$, a pair of heavy quarks at positions $\r_1, \r_2$ once such a pair had been added to the system at time $t=0$ at positions $\r_1', \r_2'$. We also define
\beq
J_Q(t; \r_1,\r_2)\equiv \chi(t,\r_2)\psi(t,\r_1),\qquad J_Q^\dagger(t; \r_1,\r_2) \equiv \psi^\dagger(t,\r_1)\chi^\dagger(t,\r_2),
\eeq
with the time dependence given by the Heisenberg representation
\beq
J_Q(t; \r_1,\r_2)={\rm e}^{iHt}\,J_Q( \r_1,\r_2)\,{\rm e}^{-iHt}.
\eeq
Here, $H$ is the full  hamiltonian, which can  generally be decomposed into three contributions:  
\beq\label{eq:hamiltoniandecomp}
H=H_Q+H_{med}+H_{int},
\eeq
where $H_Q$ is the (non relativistic) hamiltonian describing the heavy fermions in vacuum, 
$
H_{med}$ is the hamiltonian of the medium in which the $Q\bar Q$ system propagates, and 
$
H_{int}$
represents the interactions between the medium and the heavy fermions.  Various model hamiltonians will be considered in this paper.

In general, the expectation value in Eq.~(\ref{eq:2parta}) will be a thermal average, 
\beq
G^>(t,\r_1;t,\r_2|0,\r_1';0,\r_2')=\frac{1}{Z} {\rm Tr} \left\{{\rm e}^{-\beta H}
J_Q(t; \r_1,\r_2)
J_Q^\dagger(0; \r_1',\r_2')\right\},  \label{eq:2partb}
\eeq
with  $Z={\rm Tr} {\rm e}^{-\beta H}$ and $\beta$ the inverse temperature, $\beta\!=\!1/T$. If we denote by $\ket{n}$ the eigenstates of $H$, and by $E_n$ the corresponding eigenvalues, one can write for $G^>(t)$ the following formal\footnote{Formal, because we treat all states here as discrete states.} expansion (in order to alleviate the notation, we omit the spatial coordinates):
\beq
G^>(t)=\frac{1}{Z}\sum_{n}e^{-\beta E_{n}}\sum_{m}e^{i(E_{n}-E_m)t}
\bra{n}J_Q \ket{m} \langle m|J_Q^\dagger  |n\rangle.
\label{eq:formal_expansion}
\eeq
As suggested by Eq.~(\ref{eq:formal_expansion}), $G^>(t)$ is an
analytic function of the (complex) time $t$ in the strip $-\beta<\mbox{{\rm Im }}t<0$. For
 $t=-i\beta$, this function takes the value
 \beq
 G^>(t=-i\beta)=\frac{1}{Z} {\rm Tr} \left\{  J_Q\, {\rm e}^{-\beta H}\, J_Q^\dagger\right\}. \label{Gdebeta}
 \eeq
 We shall return to this expression shortly. 
 
One can introduce, together with   $G^>(t)$, a collection of related correlators (see e.g. \cite{lb}; the conventions employed in this paper are those in \cite{Blaizot:2001nr}). 
Thus, one defines
\beq
G^<(t,\r_1;t,\r_2|0,\r_1';0,\r_2')\equiv 
\langle J_Q^\dagger(0; \r_1',\r_2')J_Q(t; \r_1,\r_2)\rangle, 
\eeq
related to  $G^>(t)$ by the KMS condition:
\beq
G^<(t)=G^>(t-i\beta), 
\eeq
and the retarded propagator
\beq\label{Gretarded}
G^R(t)\equiv i\,\theta(t)\left[G^>(t)-G^<(t)\right], 
\eeq
as well as the time-ordered propagator
\beq\label{time-ordered}
G(t)\equiv i\,\theta(t)\,G^>(t)+ i\,\theta(-t)\,G^<(t),
\eeq
which is used in perturbation theory.
Note that 
\beq\label{time-ordered2}
G(t)=G^R(t)+i\,G^<(t).
\eeq

From the Fourier transforms $G^>(\omega)$ and $G^<(\omega)$ one obtains the spectral function  $\rho_G$:
\beq\label{eq:def_spec}
\rho_G(\omega)\equiv \int_{-\infty}^\infty dt\, {\rm e}^{i\omega t} \,\left[G^>(t)-G^<(t)\right]\!=\!(1-e^{-\beta\omega})G^>(\omega)\!=\!(e^{\beta\omega}-1)G^<(\omega),
\eeq
the last equalities following from the KMS condition $G^>(\omega)={\rm e}^{\beta\omega}G^<(\omega)$. One can use the spectral function to define an analytic propagator  ($z$ complex):
\beq\label{analyticprop}
\tilde G(z)=\int_{-\infty}^{+\infty}\frac{dq^0}{2\pi}\frac{\rho_G(q^0)}{q^0-z}.
\eeq
$\tilde G(z)$ is an analytic function of $z$ in the whole complex plane except the real axis. When $z$ approaches the real axis from above, $\tilde G$ goes over to the Fourier transform of the retarded propagator: $G^R(\omega)=\tilde G(\omega+i\eta)$, with $\omega$ real.
The analytic propagator is also related to the propagator that enters perturbative calculations at finite temperature in the imaginary time formalism, namely the  Matsubara propagator: $\tilde G(i\omega_n)$ where $\omega_n=2n\pi /\beta$, $n$ integer,  is a Matsubara frequency.

As to the Fourier transform of the real-time propagator, it reads (see Eqs.~(\ref{time-ordered2}) and (\ref{eq:def_spec})):
\beq\label{FTrealtimeprop}
G(\omega,\q)=\int_{-\infty}^{+\infty}\frac{dq^0}{2\pi}\frac{\rho_G(q^0,\q)}
{q^0-\omega-i\eta}+i\rho_G(\omega,\q)N(\omega)\,,
\eeq
with
\beq
N(\omega)=\frac{1}{e^{\beta \omega}-1}\,.
\eeq
 It is useful to observe that the limits $\omega\to 0$ are in general distinct for the anaytic and the real time propagators. Using the fact that the spectral density $\rho_G(q_0)$ vanishes when $q_o=0$, one can write for the real time propagator
 \beq\label{FTrealtimeprop0}
G(\omega=0,\q)=\int_{-\infty}^{+\infty}\frac{dq^0}{2\pi}\frac{\rho_G(q^0,\q)}
{q^0}+iT \lim_{\omega\to 0}\frac{\rho_G(\omega,\q)}{\omega}\,,
\eeq
while only the first term in the equation above is present in $\tilde G(\omega=0)$, as is obvious from Eq.~(\ref{analyticprop}).

The relations above are valid not only for the propagator $G$, but also for most of the propagators that we shall encounter in this paper, to within signs for fermions. Thus for instance, the heavy fermion propagator reads (dropping again the spatial coordinates to alleviate the notation)
\beq
S(t)\equiv i\,\langle {\rm T} \psi(t)\psi^\dagger(0)\rangle=i\,\theta(t) S^>(t)-i\,\theta(-t) S^<(t),
\eeq
which can be also expressed as:
\beq
S(t)=S^R(t)-iS^<(t)
\eeq
with $S^R(t)=i\theta(t) \left[  S^>(t)+S^<(t)  \right]$.

We shall, most of the time in this paper, consider the case where the mass of the heavy quarks is taken to infinity. In this limit two important simplifications occur. First, the statistical factors describing the probability to find a heavy quark in the heat bath vanish exponentially (as ${\rm e}^{-M/T}$). It follows that the states involved in the thermal averages may be considered as vacuum states for the operators $\psi$ and $\chi$, so that $S^<$ and $G^<$ vanish. In other words,
in this limit, the time ordered propagator and the retarded propagator coincide. The same situation may be imposed in the case where $M$ is finite, by treating the heavy quarks as ``test particles", that is as particles which are not part of the heat bath.  The second simplification which occurs in the infinite mass limit is that  the heavy particles do not move, so that their propagators are proportional to delta functions of the coordinates. To see that, recall that, as we just emphasized,  for  a test particle, or in the limit of a very massive fermion, the time ordered propagator is identical to the retarded propagator. In the absence of interactions one has then
$S^0(t)=S_R^{0}(t)$ with 
\beqa\label{S0ret}
S^{0}_R(t,\x)&=&\!\!\int\frac{d\omega}{2\pi}\!\int\!\frac{d\k}{(2\pi)^3}
e^{-i\omega t}
e^{i\k\cdot\x}\frac{-1}{\omega-\frac{k^2}{2M}+i\eta}=\!
i\theta(t)\!\!\int\frac{d\k}{(2\pi)^3}e^{-i\frac{k^2}{2M} t+i\k\cdot\x}
\nonumber\\
{}&=&i\theta(t)\frac{1}{(2\pi)^3}\left(\frac{2M\pi}{it}\right)^\frac{3}
{2} 
e^{i\frac{M\x^2}{2t}}.
\eeqa
In the limit $M\!\to\!\infty$, 
\beq\label{S0}
S^{0}_R(t,\x)=i\theta(t)\delta(\x),
\eeq
so that (with $G^0$ the propagator (\ref{time-ordered}) in the absence of interactions)
\beq
\lim_{M\to\infty}G^{0}(t,\r_1;t,\r_2|0,\r_1';0,\r_2')=
i\,\theta(t)\delta(\r_1-\r_1')\delta(\r_2-\r_2').
\label{eq:noninterMinf}
\eeq
As we shall see, such delta functions can be factorized also in the presence of interactions. 

The physical information that can be extracted from the correlator (\ref{eq:2parta}) depends on whether one follows the evolution in real time or in imaginary time (Euclidean formalism). We shall consider only the  large $M$ limit here. In real time, it is convenient to consider the correlator (\ref{eq:2parta}), or rather the retarded one in Eq.~(\ref{Gretarded}), from the point of view of linear response. We consider the medium initially in thermal equilibrium at temperature $T$, and perturb it at time $t=0$ with a perturbation proportional to the operator $J_Q^\dagger (\r'_1,\r'_2)+J_Q(\r'_1,\r'_2)$. That is, for times $t>0$, the hamiltionian of the full system is $H-\lambda [ J_Q^\dagger (\r'_1,\r'_2)+J_Q(\r'_1,\r'_2)]$. The expectation value of the operator $J_Q^\dagger (\r_1,\r_2)+J_Q(\r_1,\r_2)$ at time $t\!>\!0$ is then given by the retarded propagator
\beq
\langle  J^\dagger (\r_1,\r_2)+J(\r_1,\r_2)\rangle_t= \lambda  \int_{0}^{t} dt' G_R(t; \r_1,\r_2|t';\r'_1,\r'_2).
\eeq
In imaginary time, lattice calculations provide information on the correlator
\beq
G^>(\tau,\r_1;t,\r_2|0,\r_1';0,\r_2')=\langle J(\tau; \r_1,\r_2)J^\dagger(0;\r'_1,\r'_2)   \rangle
\eeq
from which one can, in principle, reconstruct the spectral function, and then real time information. When the imaginary time $\tau$ is set equal to $\beta$, the physical information that one can extract concerns essentially equilibrium properties.  Consider  Eq.~(\ref{Gdebeta}) in the limit $M\to \infty$.  It can be written as
 \beq
 G^>(t=-i\beta)=\frac{1}{Z}\sum_n \bra{n}  J_Q\, {\rm e}^{-\beta H}\, J_Q^\dagger\ket{n}, \label{Gdebetab}
 \eeq
 where $H=H_{med}+H_{int}$  but, according to the remarks made earlier, the states $\ket{n}$ can be chosen as eigenstates of the medium Hamiltonian $H_{med}$ ($H_Q\ket{n}=J_Q\ket{n}=0$). The sum over the states $\ket n$ in Eq.~(\ref{Gdebetab})  can then be interpreted as the partition function for a system of two heavy quarks at fixed positions in a medium at temperature $T$; a basis of  states for this system is indeed provided by states of the form $J_Q^\dagger\ket{n}$, the operator  $J_Q^\dagger(\r_1,\r_2)$ creating a pair of heavy quarks at positions $\r_1,\r_2$, and $\ket{n}$ being an eigenstate of $H_{med}$. Up to a product of delta functions that takes care of the normalization of such states, $\bra{n}J_Q(r_1,\r_2)J_Q^\dagger (r_1',\r_2')\ket{n}=\delta(\r_1-\r_1')\delta(\r_2-\r_2')$, this partition function is related to the free energy of the heavy quark pair in the medium. Specifically,  
 \beq\label{freeenergy0}
 G^>(t=-i\beta)= \delta(\r_1-\r_1')\delta(\r_2-\r_2') \,e^{-\beta\Delta F_{Q\overline{Q}}(\r_1-\r_2)},
\eeq
where $\Delta F_{Q\overline{Q}}(\r_1-\r_2)$ is the free energy of the heavy fermions located at positions $\r_1$ and $\r_2$, relative to the free energy of the medium in the absence of the heavy fermions, the latter being given by $-T\ln Z$. This formula  is analogous to  that established in \cite{mclerr} which gives the change in the free-energy resulting from the addition of a heavy quark-antiquark pair  in terms of the correlator of  the associated Polyakov lines.

The correlator (\ref{eq:2parta}) studied in this paper is  related to  correlators that are  directly involved in the determination of observables. In particular, it is closely related to  the following correlator
\beq
G_M^>(t,\x)\equiv
\langle J_M(t,\x)J_M^\dagger(0,\0)\rangle,
\label{eq:gbigger}
\eeq
where the operator  $J_M(t,\x)\!\equiv\!\bar{q}(t,\x)\Gamma_M q (t,\x)$ excite mesons with  quantum numbers that depend on  $\Gamma_M$. In particular, in the vector channel, such a correlator enters  the calculation of the dilepton production rate, a quantity of experimental interest,  used for instance as  a diagnosis of the the state of matter produced in the ultra-relativistic heavy-ion collisions.
The correlator in Eq.~(\ref{eq:gbigger}) is also 
the quantity which is evaluated for imaginary times in lattice simulations, and  used 
to reconstruct the mesonic spectral functions. Finally, a  correlator such as (\ref{eq:2parta}) can also be used as a starting point for a  $T$-matrix approach such as that developed in Ref.~\cite{rapp1}. 

The correlator (\ref{eq:2parta})
is not a gauge-invariant quantity, since the $Q\bar{Q}$ pair is not produced by a local current operator. Although this is in principle not a problem 
it may be preferable, when one performs approximations, to work with gauge invariant objects.
This may be achieved by introducing a modified propagator in which a
parallel transporter is introduced to connect the heavy fermions at times $t\!=\!0$ and $t$,
as was done for instance in Ref.~\cite{lai1}.
However for the purposes of the present paper, namely the study of the $Q\overline{Q}$ free-energy and the large (real) time behavior of the heavy fermion correlator, this is not needed, as we shall verify explicitly in Sec.~\ref{sec:QEDtoy}.

\section{Non relativistic setting at $T=0$}%
\label{sec:nrtoy}                     %
We start with the simple  case where the two heavy fermions propagate in vacuum and interact through a spin-independent
instantaneous potential $V$. Although we are dealing here with a problem of non relativistic quantum mechanics, we shall use a field theory formulation that will make forthcoming generalizations easier. Thus we write the hamiltonian $H_Q$ in second quantization (we set $\hbar=1$):
\begin{multline}
H_Q=\int d\x\,\psi^\dagger(\x)\left(\frac{-\nabla^2}{2M}\right)\psi(\x)
+\int d\x\,\chi^\dagger(\x)\left(\frac{-\nabla^2}{2M}\right)\chi(\x)+\\
+\int d\x d\y\,\psi^\dagger(\x)\chi^\dagger(\y)
V(\x-\y)\chi(\y)\psi(\x)\,, 
\label{eq:hamiltonian}
\end{multline}
and the correlator in Eq.~(\ref{eq:2part}) is
\beq
G^>(t,\r_1;t,\r_2|0,\r_1';0,\r_2')\equiv
\langle 0| \chi(t,\r_2)\psi(t,\r_1)
\psi^\dagger(0,\r_1')\chi^\dagger(0,\r_2')|0\rangle,\label{eq:2part}
\eeq
where $|0\rangle$ is the vacuum of the operators $\psi$ and $\chi$ ($\psi|0\rangle=\chi|0\rangle=\langle 0|\psi^\dagger=\langle 0|\chi^\dagger=0$).

From the equation of motion of the fields in the Heisenberg picture
\beq
i\partial_t\psi(t,\r_1)=
T(\r_1)\psi(t,\r_1)
+\!\int d\y\, \chi^\dagger(t,\y)V(\r_1-\y)
\chi(t,\y)\psi(t,\r_1)\,,
\eeq
and similarly for $\chi$, one deduces the 
evolution equation for $G^>$:
\beqa
&&\!\!\!\!\!\left(i\partial_t -T(\r_1)-T(\r_2)\right)G^>(t,\r_1;t,\r_2|0,\r_1';0,\r_2')\nonumber\\
&=&\int d\x\,\langle 0|\psi^\dagger(t,\x)V(\x-\r_2)
\psi(t,\x)\chi(t,\r_2)\psi(t,\r_1)\psi^\dagger(0,\r_1')
\chi^\dagger(0,\r_2')|0\rangle \nonumber\\
&+&\int d\y\,\langle 0|\chi(t,\r_2)\chi^\dagger(t,\y)V(\r_1-\y)
\chi(t,\y)\psi(t,\r_1)\psi^\dagger(0,\r_1')
\chi^\dagger(0,\r_2')|0\rangle,\nonumber\\  \label{eq:eom2part} 
\eeqa
where $T(\r)\!\equiv\!-\nabla_\r^2/2M$.
In general, this would not be a closed equation, its
RHS involving a six-point function.
However, 
the first term
in the RHS of Eq. (\ref{eq:eom2part}) does not contribute since $\bra{0} \psi^\dagger(t,\x)=0$;  the last one can be easily evaluated by employing the equal-time
anti-commutation relations 
\beq
\left\{\chi(t,\r_2),\chi^\dagger(t,\y)\right\}=\delta(\r_2-\y)\,,
\eeq
and it can be expressed entirely in terms of $G^>$. 
One is left then with the closed equation
\beq
\label{eq:schrod2part}
\left(i\partial_t -T(\r_1)-T(\r_2)-V(\r_1-\r_2)\right)
G^>(t,\r_1;t,\r_2|0,\r_1';0,\r_2')=0
\eeq
which, not unexpectedly,  has the structure of a two-particle Schr\"odinger equation with the
boundary condition
$G^>(0,\r_1;0,\r_2|0,\r_1';0,\r_2')\!=\!\delta(\r_1\!-\!\r_1')
\delta(\r_2\!-\!\r_2')$. In fact,  denoting  by $\Psi_m(\r_1,\r_2)$ the stationary states of the  Schr\"odinger equation for the two heavy particles interacting with the potential $V$, we have
\beqa\label{G>station}
G^>(t,\r_1;t,\r_2|0,\r_1';0,\r_2')&=&\sum_{m}e^{-iE_m t}
\langle0|\chi(\r_2)\psi(\r_1)|m\rangle\langle m|
\psi^\dagger(\r_1')\chi^\dagger(\r_2')|0\rangle\nonumber\\
{}&=& \sum_{m}e^{-iE_m t}\,\Psi_m(\r_1,\r_2)\Psi_m^\star(\r_1',\r_2').
\eeqa
\begin{figure}[!tp]
\begin{center}
\includegraphics[clip,width=0.65\textwidth]{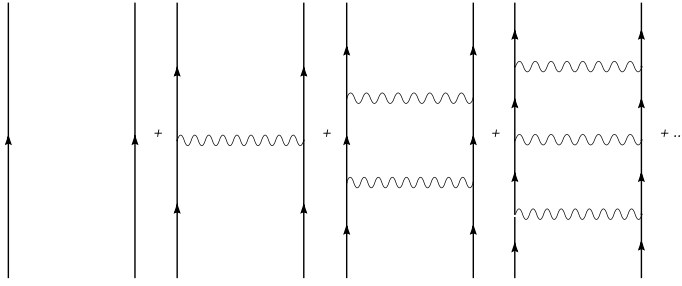}
\caption{The ladder diagrams contributing to the two-particle propagator. The wavy line denotes the  instantaneous interaction potential.}
\label{fig:ladder} 
\end{center}
\end{figure}
It is useful, in view of the forthcoming discussions, to write
the solution of Eq.~(\ref{eq:schrod2part}) as  an expansion in powers
of $V$.
In order to do so we note that the time-ordered propagator is simply $G(t)=i\theta(t)G^>(t)$ 
since $G^<(t)\!=\!0$ here. One can then get the desired expansion for $G^>(t)$  from the usual perturbative expansion for the time-ordered propagator:
\begin{multline}
-iG(t,\r_1;t,\r_2|0,\r_1';0,\r_2')=
\sum_{n=0}^\infty\frac{(-i)^n}{n!}\int_{-\infty}^{+\infty}dx_1^0\dots
\int_{-\infty}^{+\infty}dx_n^0\\
\times\bra{0} T\left( H_1(x_1^0)\dots H_1(x_n^0)
\chi(t,\r_2)\psi(t,\r_1)
\psi^\dagger(0,\r_1')
\chi^\dagger(0,\r_2')
\right)\ket{0},\label{eq:pert}
\end{multline}
where the fields are in the interaction picture and 
the interaction hamiltonian reads:
\beq
H_1(x^0)={\rm e}^{iH_0 x_0}H_1{\rm e}^{-iH_0 x_0}=\int d\x \int d^4y\,
\psi^\dagger(x)\chi^\dagger(y)
U(x-y)\chi(y)\psi(x)\,,
\eeq
with:
\beq
U(x-y)=V(\x-\y)
\delta(x^0-y^0)\,,
\eeq
and $H_0$ is the heavy quark kinetic energy.
Note that, since $\bra{0} H=H\ket{0}=0$, the various contributions to the series in Eq.~(\ref{eq:pert}) vanish unless all time arguments $x^0_i$ are between 0 and $t$. These contributions correspond to the ladder diagrams depicted in  Fig. \ref{fig:ladder}. 
The first term in the expansion is  simply the product of two
non-interacting  single-particle propagators
\beq
-iG^{0}(t,\r_1;t,\r_2|0,\r_1';0,\r_2')=
-S^{0}(t,\r_1-\r_1')\,
S^{0}(t,\r_2-\r_2')\,.\label{eq:lo2part}
\eeq

In the limit $M\!\to\!\infty$, 
 it is easy to resum the series of ladder diagrams in Eq.~(\ref{eq:pert})  in a compact form: the intermediate propagators are proportional to delta functions of the positions of the particles (see Eq.~(\ref{S0})), and the integrations over the time variables can then be trivially done. The resulting series reduces to that of an exponential.  One gets:
\beq\label{eq:nrexp00}
G(t,\r_1;t,\r_2|0,\r_1';0,\r_2')\equiv G^{0}(t,\r_1;t,\r_2|0,\r_1';0,\r_2')
\,\overline{G}(t,\r_1-\r_2),
\eeq
where $G^0$ is given by Eq.~(\ref{eq:noninterMinf}) and 
\beq
\overline{G}(t,\r_1\!-\!\r_2)=\exp\!\left[-it V(\r_1\!-\!\r_2)\right].\label{eq:nrexp0}
\eeq
(Note that $\overline{G}$ is essentially $G^>$: $G^>(t,\r_1;t,\r_2|0,\r_1';0,\r_2')=\delta(\r_1-\r'_1)\delta(\r_2-\r'_2) \overline{G}(t,\r_1\!-\!\r_2)$.) This result can of course be obtained directly from Eq.~(\ref{eq:schrod2part}), noting that in the infinite mass limit, the kinetic energy vanishes so that the coordinates $\r_1, \r_2$ play the role of parameters in the differential equation. 

One can also perform easily the analytic continuation to imaginary time and get the Euclidean correlator. In particular, setting $it=\beta$ in Eq.~(\ref{eq:nrexp0}), one obtains that the change in the free energy is independent of the temperature and  equal to the potential $V(\r_1-\r_2)$ (see Eq.~(\ref{freeenergy0})). Note that the same analytic continuation performed {\em before}  taking the infinite mass limit (i.e., setting $t=-i\beta$ in Eq.~(\ref{G>station})) yields 
\beq
G^>(t=-i\beta,\r_1;-i\beta,\r_2|0,\r_1';0,\r_2')=\sum_{m} e^{-\beta E_m }\Psi_m(\r_1,\r_2)\Psi_m^\star(\r_1',\r_2').
\eeq
In particular, for $\r_1=\r_1'$ and $\r_2=\r_2'$,
\beq
G^>(-i\beta,\r_1;-i\beta,\r_2|0,\r_1';0,\r_2')=\sum_{m} e^{-\beta E_m }|\Psi_m(\r_1,\r_2)|^2
\eeq
is proportional to the probability to find the two heavy particles at positions $\r_1,$ $\r_2$ when these are  in thermal equilibrium at temperature $1/\beta$.

We turn now to the case where the heavy particles propagate in a medium that they can polarize. We model this medium by a set of  of light fermions of mass $m$, described by the field
$\phi$. These light fermions interact with the heavy particles with the interaction potential $V$. 
The hamiltonian takes then the form in Eq.~(\ref{eq:hamiltoniandecomp})
with $H_Q$ defined in Eq. (\ref{eq:hamiltonian}), 
\beq
H_{med}=\int d\x\,\phi^\dagger(\x)\left(\frac{-\nabla^2}{2m}\right)\phi(\x)-\frac{1}{2}\int\!d\x d\y\,\phi^\dagger(\x)\phi^\dagger(\y)
V(\x-\y)\phi(\y)\phi(\x),
\eeq
and
\begin{multline}
H_{int}=-\int\! d\x d\y\,\psi^\dagger(\x)\phi^\dagger(\y)
V(\x-\y)\phi(\y)\psi(\x)\\
+\int\!d\x d\y\,\chi^\dagger(\x)\phi^\dagger(\y)
V(\x-\y)\phi(\y)\chi(\x).
\end{multline}
The correlator of the heavy fermions is still  given by Eq.~(\ref{eq:2part}) with now $\ket{0}$ the ground state of the many body system, which however still plays the role of the vacuum for the heavy fermions. 

The exact evolution equation for $G^>$ 
reads now:
\begin{multline}
\left(i\partial_t\!-\!T_1\!-\!T_2\!-\!V(\r_1\!-\!\r_2)\right)
G^>(t,\r_1;t,\r_2|0,\r_1';0,\r_2')=\\
\int d\x \bra{0}\phi^\dagger(t,\x)V(\x-\r_2)
\phi(t,\x)\chi(t,\r_2)\psi(t,\r_1)\psi^\dagger(0,\r_1')
\chi^\dagger(0,\r_2')\ket{0}\\
-\int d\y\bra{0}\chi(t,\r_2)\phi^\dagger(t,\y)V(\r_1-\y)
\phi(t,\y)\psi(t,\r_1)\psi^\dagger(0,\r_1')
\chi^\dagger(0,\r_2')\ket{0}\,.
\end{multline}
This equation reveals the full complexity of the many-body problem that we are dealing with. It cannot be solved exactly, nor in general be reduced to a simple Schr\"odinger equation, or a closed equation for $G^>$. However the structure of the corresponding  time-ordered  propagator can be easily analyzed in perturbation theory, as we did before. 
\begin{figure}[!tp]
\begin{center}
\includegraphics[clip,width=0.7\textwidth]{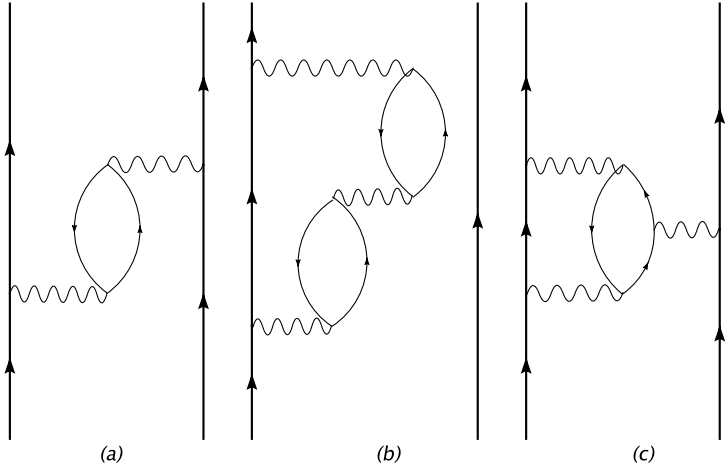}
\caption{Some of the processes occuring when a
of a $Q\overline{Q}$ pair propagates in a medium of light fermions.
The diagrams (a) and (b) are accounted for by the effective interaction.
On the contrary diagrams (c) is not included in the effective interaction.}
\label{fig:medium} 
\end{center}
\end{figure}
Typical diagrams are displayed in Fig.~\ref{fig:medium}, where several types of processes are exhibited: self-energy corrections, and various modifications of the effective interaction between the heavy fermions. In order to progress in our analysis, we shall assume, as done in many studies, that the dominant 
effect of the medium is to modify the interaction potential through screening
effects. 
\begin{figure}[!tp]
\begin{center}
\includegraphics[clip,width=0.7\textwidth]{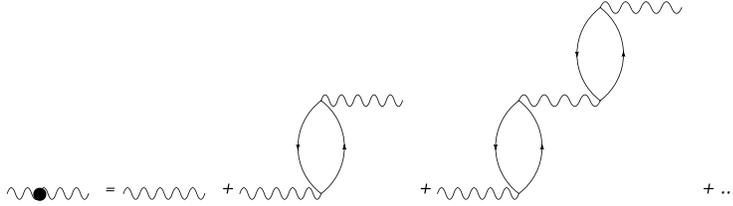}
\caption{The diagrams contributing to the effective interaction.}
\label{fig:ring} 
\end{center}
\end{figure}
We can then write a \emph{resummed} perturbative expansion by 
employing the following \emph{effective}
interaction hamiltonian (in the interaction picture):
\begin{multline}
H_1^{\rm eff}(x_0)=\int d\x \int d^4y\,\psi^\dagger(x)\chi^\dagger(y)
U^{Q\bar{Q}}(x-y)
\chi(y)\psi(x)+\\
+\frac{1}{2}\int d\x \int d^4y\,\psi^\dagger(x)\psi^\dagger(y)U^{QQ}(x-y)
\psi(y)\psi(x)+\\
+\frac{1}{2}\int d\x \int d^4y\,\chi^\dagger(x)\chi^\dagger(y)U^{\bar{Q}
\bar{Q}}(x-y)
\chi(y)\chi(x)\,.\label{eq:effintham}
\end{multline}
The diagrams contributing to the 
effective $Q\bar{Q}$ and  $QQ$
(and $\bar{Q}\,\bar{Q}$) interactions are displayed in Fig.~\ref{fig:ring}. The effective interactions are conveniently expressed in terms of their Fourier transfoms:
\beqa\label{eq:medium1}
U^{Q\bar{Q}}(\omega, \q)&\equiv&
\frac{U_0(\omega, \q)}{1-\Pi(\omega,\q)U_0(\omega,\q)}=\frac{V(\q)}{1-\Pi(\omega,\q)V(\q)}, \nonumber\\
U^{QQ}(\omega,\q)&\equiv&
\frac{U_0(\omega,\q)\Pi(\omega,\q)U_0(\omega,\q)}{1-\Pi(\omega,\q)U_0(\omega,\q)}=U^{\bar{Q} \bar{Q}},
\eeqa
where
$\Pi(\omega,\q)$ is the particle-hole bubble. Note that the $\omega$-dependence of $\Pi$ makes the medium induced interaction no longer
instantaneous. 
 
\begin{figure}[!tp]
\begin{center}
\includegraphics[clip,width=0.25\textwidth]{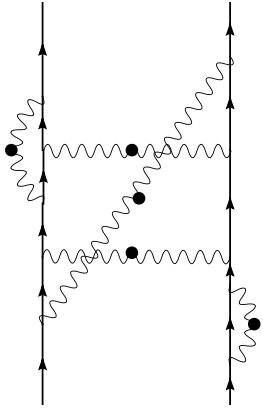}
\caption{An example of the diagrams included in the exact exponentiation
occurring in the $M\!\to\!\infty$ limit. The wavy lines with  black dots denote the effective interaction (see Fig.~\ref{fig:ring}).}
\label{fig:expo} 
\end{center}
\end{figure}

We can now study how the medium-modified interaction affects the two-particle
correlator, though being aware that this does not exhaust all the possible
processes occurring during the in-medium propagation.
Note that in the case of a long-range interaction, like in the electron-gas
problem, this choice is sufficient to account for most medium effects \cite{wal}.  However, our purpose here being only to illustrate some specific effects, no further effort will be made to justify keeping only this particular class of diagrams. In the same spirit we shall ignore tadpole diagrams, assuming that the medium has the necessary properties to ensure their cancellation. A typical diagram contributing to the two-particle  propagator is then given in
Fig. \ref{fig:expo}.

The  treatment of the effective interaction  remains a difficult task in general, but can be done exactly 
in the  limit $M\!\to\!\infty$.
One starts from the perturbative expansion given in Eq.~(\ref{eq:pert}) with the effective interaction hamiltonian introduced in Eq.~(\ref{eq:effintham}).
In the $M\!\to\!\infty$ limit
the various contractions that involve a given heavy fermion contibute an overall factor $\theta(t) \delta(\r-\r')$. Hence, in any order,  one can factorize $\theta(t)\delta(\r_1-\r_1')\delta(\r_2-\r_2')$
corresponding to the non-interacting result given in Eq.
(\ref{eq:noninterMinf}).
The different ways of contracting the fields simply amount to the different
temporal orderings of the interaction vertices $x_i^0$ and $y_i^0$.
The integrations over these variables can be done freely and lead  again to an exponentiation. The final result takes the form of Eq.~(\ref{eq:nrexp00}) with here
\beq
\overline{G}(t,\r_1\!-\!\r_2)=\exp\!\left[-i\!\int_0^t dx^0\!\int_0^t
dy^0\left(U^{Q\bar{Q}}(x^0\!-\!y^0,\r_1\!-\!\r_2)\!+\!
U^{QQ}(x^0\!-\!y^0,\0)\right) \right].\label{eq:nrexp}
\eeq
By expressing the effective interaction in terms of its Fourier transform, one obtains:
\begin{multline}
\overline{G}(t,\r_1-\r_2)\!=\!
\exp\!\Big[-2i\!\int\frac{d\omega}{2\pi}\int\frac{d\q}{(2\pi)^3}
\frac{1-\cos(\omega t)}{\omega^2}\times\\
\times\left(U^{QQ}(\omega,\q)
+e^{i\q\cdot(\r_1-\r_2)}U^{Q\overline{Q}}(\omega,\q)\right)\!
\Big]. \label{eq:2corrnr}
\end{multline}
The large time behavior of this expression is easily obtained by using 
\beq
\lim_{t\to\infty}\frac{1-\cos(\omega t)}{\omega^2}=\pi t \delta(\omega).
\label{eq:limcos}
\eeq
One gets:
\beq
\overline{G}(t,\r_1-\r_2)\!\underset{t\to\infty}{\sim} 
\exp\left(  -it V_\infty(\r_1-\r_2) \right),\label{eq:2nrtinfty}
\eeq
where
\beq\label{Vinfty}
V_\infty(\r_1-\r_2) =\!\int\frac{d\q}{(2\pi)^3}\left(U^{QQ}(\omega\!=\!0,\q)
\!+\!e^{i\q\cdot(\r_1-\r_2)}U^{Q\overline{Q}}(\omega\!=\!0,\q)
\right).
\eeq
 Quite remarkably, aside from the self-energy corrections, this large time behavior (\ref{eq:2nrtinfty}) is the same result as would have been obtained with an instantaneous potential  $V_\infty(\r_1-\r_2) =U^{Q\bar Q}(\omega\!=\!0,\r_1-\r_2)$ (compare with Eq.~(\ref{eq:nrexp0}) that expresses the resummation of ladder diagrams involving the instantaneous potential $V(\r_1-\r_2)$).
We are facing here an issue closely related to the problem of recovering a
potential model from a two-body relativistic equation (with the energy
dependence of the interaction arising from the propagator of the exchanged meson). Also in this case, in order to get, in the large mass limit, an energy-independent interaction kernel, one has to sum all possible ladder and crossed diagrams~\cite{gross}.
Note that the  evolution equation obeyed by  $\overline G(t)$ at large time is simply:
\beq
\lim_{t\to\infty}\left[i\partial_t-\!\int\frac{d\q}{(2\pi)^3}\!\left(U^{QQ}
(\omega\!=\!0,\q)
\!+\!e^{i\q\cdot(\r_1-\r_2)}U^{Q\overline{Q}}(\omega\!=\!0,\q)\right)\right]
\overline{G}(t,\r_1-\r_2)\!=0.
\eeq
This is the analog of Eq. (4.2) in \cite{lai1}. 

So far, we have been working in real time at $T=0$, considering the propagation of the $Q\overline{Q}$ pair in a cold-dense medium of light particles. However, one can easily extend the previous calculations to address the case where the heavy pair propagates in a medium at
finite temperature. This can be done by exploiting the analyticity properties recalled in  Sec.~\ref{sec:basic}, or by repeating the calculation above in the imaginary time formalism.  In fact, when it is carried in the imaginary time formalism, the preceding perturbative analysis goes through unchanged except for the replacements $t\to -i\tau$, $-iU(t=-i\tau)\to {\cal U}(\tau)$ (in Fourier space, the relation between $U(\omega)$ and ${\cal U}(\omega)$ is the same as that between, respectively,  the real time and the analytic propagators discussed in  Sec.~\ref{sec:basic}).  In the infinite mass case, the exponentiation of the interaction effects still holds, and one gets
\begin{multline}
\overline{G}(-i\tau,\r_1\!-\!\r_2)\\ =\exp\!\left[-\!\int_0^\tau d\tau'\!\int_0^\tau d\tau''\,\left(  {\cal U}^{Q\bar{Q}}(\tau'\!-\!\tau'',\r_1\!-\!\r_2)\!+\!{\cal U}^{QQ}(\tau'\!-\!\tau'',\r_1\!-\!\r_2)\!\right)\right].\label{eq:nrexp2a}
\end{multline}
At this point, it is convenient to express the interaction in Fourier space, for instance:
\beq
{\cal U}^{Q\bar{Q}}(\tau'\!-\!\tau'',\r_1\!-\!\r_2)=\frac{1}{\beta}\sum_{n}e^{-i\omega_n(\tau'-\tau'')}\!\int\frac{d\q}{(2\pi)^3}e^{i\q\cdot(\r_1-\r_2)}
{\cal U}^{Q\bar{Q}}(i\omega_n,\q)\,
\eeq
where $\omega_n=2n\pi T$ is a Matsubara frequency. 
The integals over $\tau'$ and $\tau ''$ in Eq.~(\ref{eq:nrexp2a}) can then be easily done.  In particular, for $\tau\!=\!\beta$, one gets
\beq
\overline{G}(-i\beta,\r_1\!-\!\r_2)=\exp\!\left[-\beta\int\frac{d\q}{(2\pi)^3}e^{i\q\cdot(\r_1-\r_2)}\left({\cal U}^{Q\bar{Q}}(\omega=0,\q)+{\cal U}^{Q{Q}}(\omega=0,\q)\right)\right],\label{eq:nrexp2b}
\eeq
allowing us  to identify the free-energy as 
\beq\label{freenonrel}
\Delta F_{Q\bar Q}=\int\frac{d\q}{(2\pi)^3}e^{i\q\cdot(\r_1-\r_2)}\,\left( {\cal U}^{Q\bar{Q}}(\omega=0,\q)+{\cal U}^{Q{Q}}(\omega=0,\q)\right).
\eeq
Note that, in spite of the similarity between Eqs.~(\ref{freenonrel}) and (\ref{Vinfty}),   there is a difference between the two expressions, of the same nature as the difference between the static limits of the real time and the analytic propagator that has been already emphasized in Sec.~\ref{sec:basic}. Eq.~(\ref{freenonrel}) corresponds to the static limit of the analytic propagator and is always real, as appropriate for a free energy. In contrast, $V_\infty$, given by Eq.~(\ref{Vinfty}), may contain an imaginary part. An explicit example will be provided in Sec.~\ref{sec:QEDtoy}.

We have addressed in this section some limiting cases in which it was possible to get
a \emph{closed}, Schr\"odinger-like, evolution equation for the two-particle
correlator: 
the trivial case of in-vacuum propagation and instantaneous interaction; 
the case of a retarded in-medium interaction between infinitely-heavy
fermions. In the latter case, the result follows from an exponentiation property that holds in the limit of a large mass for the heavy fermions, and asymptotically for large times. The following section will clarify the origin of the exponentiation by using an explicit  model for the interaction, based on a simple meson exchange.

\section{Interactions due to meson-exchange}%
\label{sec:scalartoy}         %

\begin{figure}[!tp]
\begin{center}
\includegraphics[clip,width=0.9\textwidth]{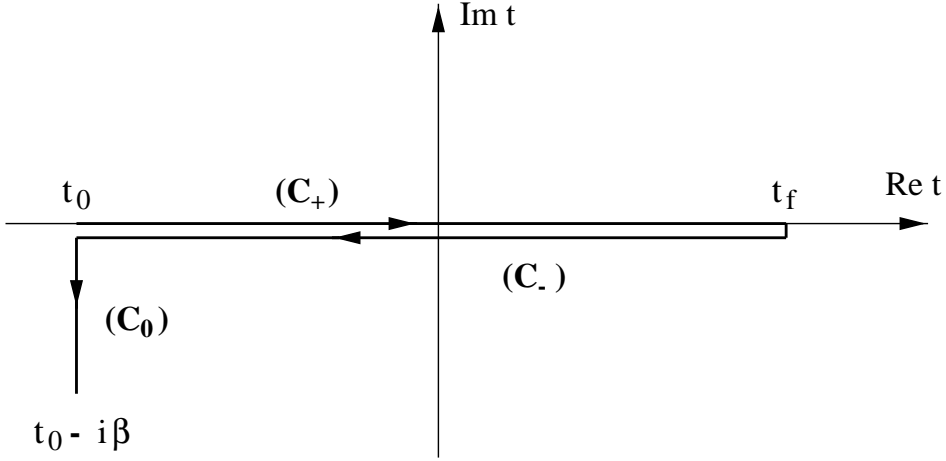}
\caption{The complex-time contour $C\equiv C_+\cup C_-\cup C_0$
used in the calculation of the real time  $Q\overline{Q}$ correlator at finite temperature. }
\label{fig:cont} 
\end{center}
\end{figure}

We now 
consider a model in which the heavy fermions propagate in a thermal bath
of scalar mesons
of mass $m$, described by the field $\Phi$,
to which they are linearly coupled.
The corresponding hamiltonian reads:
\begin{multline}
H\!=\!\int d\x\,\psi^\dagger(\x)\!\left(\frac{-\nabla^2}{2M}\!+g\Phi(\x)
\!\right)\!\psi(\x)
+\!\int d\x\,\chi^\dagger(\x)\!\left(\frac{-\nabla^2}{2M}\!+g\Phi(\x)\right)
\!\chi(\x)+\\
+\int d\x\left[\frac{1}{2}\pi^2(\x)+\frac{1}{2}(\nab\Phi(\x))^2+
\frac{1}{2}m^2\Phi^2(\x)\right],
\label{eq:mesonham}
\end{multline}
where $\pi(x)$ is the conjugate momentum of the field $\Phi(x)$.
This hamiltonian is of the same form as in Eq.~(\ref{eq:hamiltoniandecomp}), with $H_Q$ the  kinetic energy of the heavy fermions, $H_{med}$ the hamiltonian of free scalar mesons, and $H_{int}$ the linear coupling between the fermions and the scalar field. Note, however, that the dynamics of the two heavy fermions in vacuum, involves the full hamitlonian: in vacuum, as well as at finite temperature, the heavy fermions interact with each other through meson exchange.

We want to evaluate the correlator
\beq
G^>(t,\r_1;t,\r_2|0,\r_1';0,\r_2')=\frac{1}{Z} {\rm Tr} \left\{{\rm e}^{-\beta H}
\chi(t,\r_2)\psi(t,\r_1)
\psi^\dagger(0,\r_1')\chi^\dagger(0,\r_2')\right\},  \label{eq:2partb2}
\eeq
with $Z={\rm Tr} {\rm e}^{-\beta H} $. This can be done either by working in imaginary time, and then  perform an analytical continuation to real values of $t$ by exploiting
the analyticity properties of $G^>(t)$ or by using Keldish-Schwinger countour techniques, such as the one  depicted in Fig.~\ref{fig:cont}.  We shall use the latter in what follows, returning to the analytic continuation towards the end of this section. 

The formalism used in this section, and the following one, follows closely that  employed in Ref.~\cite{BNbla} for the discussion of the damping rate of a hard fermion.
Our strategy to calculate  $G^>$  will rely on the following expression for the (time-ordered) propagator:
\beq\label{basicformprop}
G(t,\r_1;t,\r_2|0,\r_1';0,\r_2')=Z^{-1}\int[{\cal D}\Phi]G_\Phi(t,\r_1;t,
\r_2|0,\r_1';0,\r_2')\,e^{iS_C[\Phi]}\,
\eeq
where 
$G_\Phi$ is the pair propagator in a given background field
$\Phi(\x,t)$ and 
\beq
Z\equiv\int[{\cal D}\Phi]e^{iS_C[\Phi]}.
\eeq
The propagator $G_\Phi(t,\r_1;t,
\r_2|0,\r_1';0,\r_2')$ is simply the product of the two propagators of the heavy fermions in the external field $\Phi$: 
\beq
iG_\Phi(t,\r_1;t,\r_2|0,\r_1';0,\r_2')= S_\Phi(t,\r_1-\r'_1)\,
S_\Phi(t, \r_2-\r'_2).
\eeq
The  average over the configurations  of $\Phi$, repesented by the functional integral in Eq.~(\ref{basicformprop}), can be done easily since the action is quadratic:
\beq\label{eq:effaction}
S_C[\Phi]=-\frac{1}{2}\int_C d^4x\int_C d^4y\,\Phi(x)D_C^{-1}(x-y)\Phi(y).
\eeq
The field $\Phi$ is defined on the contour $C$ in the
complex-time plane displayed in Fig.
\ref{fig:cont} and obeys the periodicity condition $\Phi(t)\!=\!
\Phi(t-i\beta)$. The corresponding propagator along the $C$ contour satisfies the KMS conditions and is given by:
\beq
-iD_{C}(t)\equiv \theta_C(t)D^>(t)
+\theta_C(-t)D^<(t),
\eeq
In the above, in order to give an unambiguous meaning to $\theta_C(t)$, one
implicitly assumes the path in the complex-time plane being parametrized by a real parameter $u$,  $t\!=\!t(u)$, with $u$ increasing as $t$ runs along the contour. Note that the
propagator along $C_+$ coincides with the real-time propagator defined in Eq. (\ref{time-ordered}), while the one along $C_0$ is closely related to the
imaginary-time Matsubara propagator, namely
\beq
-iD_{C_0}(t\!=\!-i\tau)\equiv\Delta(\tau).\label{eq:linkC_0} 
\eeq

The usefulness of Eq.~(\ref{basicformprop}) relies on the possibility to calculate explicitly  $G_\Phi$, which can be done easily in the  limit  $M\to\infty$, to which we shall restrict ourselves here. In this limit  the propagator $S_\Phi$ coincides with the retarded propagator (note that in Eq.~(\ref{basicformprop})  $S_\Phi$, or $G_\Phi$,  satisfies the same boundary conditions as $G$, namely retarded conditions; these differ from the KMS conditions satisfied by the meson propagator). We have:
\beq
S_\Phi(t, \r_i-\r'_i)=i\theta(t)\delta(\r_i-\r'_i)\exp\left(-ig\int_0^t dt'
\Phi(t',\r_i)\right), 
\eeq
so that 
\beq
G_\Phi(t,\r_1;t,\r_2|0,\r_1';0,\r_2')=i\theta(t)
\delta(\r_1\!-\!\r_1')\delta(\r_2\!-\!\r_2')
\overline{G}_\Phi(t,\r_1-\r_2)
,\label{eq:gphi}
\eeq
where 
\beq
\overline{G}_\Phi(t,\r_1-\r_2)=\exp \left(\!-i\!\int d^4z
J(z)\Phi(z)
\right),
\eeq
and
\beq
J(z)
=g\int_{0}^t dt'\delta(z^0-t')[\delta(\z-\r_1)+\delta(\z-\r_2)]\,.
\label{eq:j12def}
\eeq
The average over   the  configurations of the field $\Phi$ can now be performed, using Eq.~(\ref{basicformprop}). One gets
\beq
\overline{G}(t,\r_1-\r_2)=\exp\left[\,\frac{i}{2}\int_{C_+} d^4x\int_{C_+} d^4y
\,J(x)D(x-y)J(y)\right]\,,
\label{eq:2partBNjj}
\eeq
where only the
$C_+$ part of the contour contributes, and accordingly $D$ is the real-time
propagator of the scalar meson.
Inserting the current (\ref{eq:j12def}) into Eq.
(\ref{eq:2partBNjj}) one gets:
\beq
\overline{G}(t,\r_1-\r_2)=
\exp\left[\,i\,g^2\int_0^t ds \int_0^t ds'\left(D(s\!-\!s',\0)\!+\!
D(s\!-\!s',\r_1\!-\!\r_2)\right)\right]\,,\label{eq:2scal}
\eeq
which is analogous to Eq. (\ref{eq:nrexp}) obtained in the
previous section.
Everything follows then in close analogy with what was found in the non-relativistic
model of Sec. \ref{sec:nrtoy}, with the meson propagator playing now the role of the effective interaction.  
In particular the large-time behavior of the $Q\overline{Q}$ propagator
is given by Eq.~(\ref{eq:2nrtinfty}) in terms of the static propagator  which, at his stage,  is  independent of the temperature: 
\beqa 
V_\infty(\r_1-\r_2)&=&-g^2\int\frac{d\q}{(2\pi)^3}\left(1+e^{i\q\cdot
(\r_1-\r_2)}\right) D(\omega=0,\q)\nonumber\\&=&-g^2\int\frac{d\q}{(2\pi)^3}\left(1+e^{i\q\cdot
(\r_1-\r_2)}\right)\frac{1}{\q^2+m^2}. \label{Vinftyscalar}
\eeqa
The effective potential is  attractive, as expected for  scalar exchange. The first term, which represent a self-energy correction, can be evaluated in dimensional regulatization
(as done e.g. in Ref.~\cite{lai1}),
while the second one yields a Yukawa potential.
One gets (with $r\!\equiv\!|\r_1-\r_2|$)
\beq\label{eq:larget_scalar}
V_\infty(\r_1-\r_2)= \frac{g^2}{4\pi}\left(m-\frac{e^{-mr}}{r}\right)\,.
\eeq

We now allow the scalar meson to couple with massless fermions 
through the same Yukawa coupling. 
The thermal bath is then composed of scalar mesons described by the field 
$\Phi$ and massless fermions. On the other hand 
the heavy fermions are still treated as test particles.
Under the hypothesis that these heavy fermions  
interact  mainly with the soft (i.e., carrying momenta $q_0,|\q|\ll T$) modes 
of the scalar field, the effective action for the averaging over the $\Phi$- configurations will be given by the ``hard thermal loop'' (HTL) effective action of the soft modes (see the corresponding discussion in Ref.~\cite{BNbla}).
This is simply given by the same gausiann expression as in Eq.~(\ref{eq:effaction}), but with $D_C$ embodying now also self-energy corrections accounting for
the effect of the light fermions on the meson propagation. There is of course a close analogy with the particle-hole bubble summation of the previous section. 
In the HTL approximation the meson self-energy (see Appendix 
\ref{app:scalarHTL}) is momentum-independent and  provides a mere correction to the mass of the meson, which becomes
\beq
 m_D^2(T)=m^2+\Pi=m^2+N_{\rm dof}\frac{g^2T^2}{6}\,,
\eeq
with the factor $N_{\rm dof}$ counting the  internal degrees of
freedom of the light fermions other than spin (flavor, color...).
This increase of the mass  corresponds, in coordinate space,  to the screening of the interaction potential and $m_D$ is commonly referred to as the Debye screening mass.

We turn now to the Euclidean propagator, where the heavy-quark current is defined along the $C_0$ part of the contour. 
Eq.~(\ref{eq:2scal}) becomes then
\beq
\overline{G}(-i\tau,\r_1-\r_2)
=\exp\left[g^2\int_0^\tau d\tau'\!\int_0^\tau d\tau''
\!\int\frac{d\q}{(2\pi)^3}\left(1\!+\!e^{i\q\cdot(\r_1-\r_2)}\right)
\!\Delta(\tau'-\tau'',\q) \right],\label{eq:exponentiation}
\eeq
where the imaginary-time propagator is defined in Eq.~(\ref{eq:linkC_0}). By expressing the latter in Fourier space
\beq
\Delta(\tau'\!-\!\tau'',\q)=\frac{1}{\beta}\sum_{n}e^{-i\omega_n(\tau'-\tau'')}\Delta(i\omega_n,\q)\, ,
\eeq
one can easily performs the imaginary time integrals and get
\begin{multline}
\overline{G}(-i\tau,\r_1-\r_2)
=\exp\left[   g^2
\!\int\frac{d\q}{(2\pi)^3}\left(  1\!+\!e^{i\q\cdot(\r_1-\r_2)}     \right)\right.\times \\
\times\left. \left(    \frac{\tau^2}{\beta} \Delta(i\omega_n=0,\q)+\frac{2}{\beta} \sum_{n\ne 0}(1-\cos \omega_n\tau)\Delta(i\omega_n,\q)      \right)
 \right],\label{eq:exponentiation2a}
\end{multline}
where $\omega_n=2n\pi T$ is a Matsubara frequency. In particular, for $\tau=\beta$, the cosine does not contribute and one has simply
\beq\label{analprop2}
\overline{G}(-i\beta,\r_1-\r_2)=\exp\left[\beta g^2\!\int\frac{d\q}{(2\pi)^3}
\left(1\!+\!e^{i\q\cdot(\r_1-\r_2)}\right)\int\frac{dq^0}{2\pi}
\frac{\rho_D(q^0,\q)}{q^0}\right],
\eeq
where we have used the expression (\ref{analyticprop}) in order to express the Matsubara propagator $\Delta(i\omega_n=0)$ in terms of the spectral function $\rho_D$ of the meson propagator. 

Note that although the expression (\ref{analprop2}) looks like the analytic continuation of the expression (\ref{eq:2nrtinfty}) (with the corresponding $V_\infty$), much information contained in the Euclidean propagator (\ref{eq:2imagtau}) is lost when we specify $\tau=\beta$, because of the cancellation of the terms involving non vanishing Matsubara frequencies (the cosine term in Eq.~(\ref{eq:exponentiation2a})).   In the expression (\ref{analprop2}), the frequency integral is the static limit of the (analytic) propagator, which coincides here with the static limit of the real time propagator (this is not so in general, as the example in the next section will show). 

Further insight on this issue can be gained by proceeding slighlty differently so as to make the analytic continuation to real time straightforward. Let us then express the Matsubara propagator in Eq.~(\ref{eq:exponentiation2a})  in terms of the spectral function $\rho_D$ as follows \beq
\Delta(\tau'-\tau'', \q)=\int\frac{dq_0}{2\pi}e^{-iq_0(\tau'-\tau'')}\rho_D(q_0,\q)
[\theta(\tau'-\tau'')+N(q^0)].
\label{eq:realPhi}
\eeq
This also allows us to perform the integrations over $\tau'$ and $\tau''$, with the result \begin{multline}
\overline{G}(-i\tau,\r_1-\r_2)=\exp\left\{g^2\!\int\frac{d\q}{(2\pi)^3}
\left(1\!+\!e^{i\q\cdot(\r_1-\r_2)}\right)\int\frac{dq^0}{2\pi}
\frac{\rho_D(q^0,\q)}{q^0}\tau\right\}\times\\
\times\exp\left\{g^2\!\int\frac{d\q}{(2\pi)^3}
\left(1\!+\!e^{i\q\cdot(\r_1-\r_2)}\right)\int\frac{dq^0}{2\pi}
\frac{\rho_D(q^0,\q)}{(q^0)^2}(e^{-q^0\tau}-1)(1+N(q^0))\right\}\times\\
\times\exp\left\{g^2\!\int\frac{d\q}{(2\pi)^3}
\left(1\!+\!e^{i\q\cdot(\r_1-\r_2)}\right)\int\frac{dq^0}{2\pi}
\frac{\rho_D(q^0,\q)}{(q^0)^2}(e^{q^0\tau}-1)N(q^0)\right\}\,.\label{eq:2imagtau}
\end{multline}
The same expression, with $\tau $ replaced by $it$, holds for $\overline{G}(t,\r_1-\r_2)$, as expected from the analyticity of $\overline{G}(-i\tau,\r_1-\r_2)$ for $\tau<\beta$. One can then rearrange the terms in order to recover the results obtained earlier, e.g.  Eq.~(\ref{Vinftyscalar}). However,  in the limit $\tau\!=\!\beta$ the last
two exponentials in Eq. (\ref{eq:2imagtau}) cancel each other and one is left
 with Eq.~(\ref{analprop2}).

It follows from Eq.~(\ref{analprop2})  that the free energy  of the $Q\bar Q$ pair  is given by  
\beq
\Delta F_{Q\overline{Q}}(r,T)=V_\infty(r)
=\frac{g^2m_D}{4\pi}-\frac{g^2}{4\pi}\frac{e^{-m_D r}}{r}\,.
\eeq
Thus in this simple model the free-energy that is obtained from the imaginary-time two-particle propagator
at $\tau\!=\!\beta$,
can be identified with the potential to be used to study the real-time propagation of a $Q\overline{Q}$ pair.
In the next section we shall find that extra contibutions to the potential may arise, that are not captured by an imaginary-time equilibrium calculation at $\tau\!=\!\beta$.
%
\section{Heavy fermions in a hot QED plasma}%
\label{sec:QEDtoy}             %
We now discuss the case, closer to hot QCD, in
which the $Q\bar {Q}$ pair is placed in a thermal bath of photons and
light fermions. We consider only the limit of infinite mass for the heavy fermions, and, to be specific, we choose the Coulomb gauge to describe the electromagnetic field (our final results are independent of the choice of the gauge). The 
 hamiltonian describing the
system reads:
\beq
H=g \int d\x\, A_0(\x)\left(  -\psi^\dagger (\x)  \psi (x)
+  \chi^\dagger (x) \chi(x)\right) +H_{\rm Cb}+H
_{\rm f}
\eeq
with $H_{\rm Cb}$ the photon hamiltonian in the Coulomb gauge, and $H
_{\rm f}$
the hamiltonian of the light fermions, including their interaction with the electromagnetic field. 

\begin{figure}[!tp]
\begin{center}
\includegraphics[clip,width=0.7\textwidth]{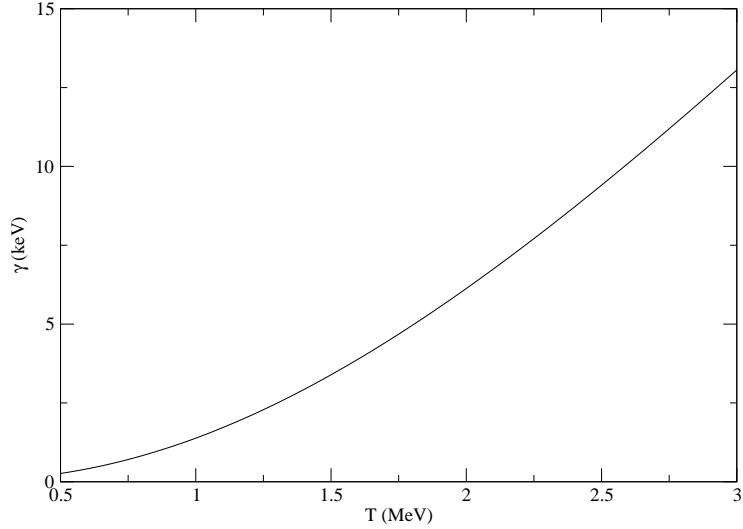}
\caption{The damping factor for a heavy fermion-antifermion pair placed in a
hot QED plasma (of photons, electrons and positrons) as a function of the
temperature. The distance between the heavy fermions has been set equal to the
mean radius of the ground state of a $\mu^+\mu^-$ atom.}\label{fig:gamma} 
\end{center}
\end{figure}

In order to evaluate the real-time
propagator of Eq. (\ref{eq:2parta})
we proceed as in the previous Section and consider first the propagation of
the heavy fermions in a
given background field $A_\mu$.
Then we average over all the possible configurations
of the gauge field. 
The treatment given here is similar to that of the fermion damping 
rate presented in Ref.~\cite{BNbla}.  One gets, using the notation of Eq. (\ref{eq:gphi}):
\beq
\overline{G}(t,\r_1-\r_2)=
\langle\exp\left(-i\int d^4z J_0(z)A_0(z)\right)
\rangle\,,
\eeq
with
\beq
J_0(z)\equiv \int_{0}^t dt' \delta(z^0-t')
\left[-g\delta(\z-\r_1)
+g \delta(\z-\r_2)\right]\,.\label{eq:currQED}
\eeq
Assumimg that the heavy fermions interact mostly with the soft modes of the
electromagnetc field, we use the HTL effective action
in order to perform the average over the gauge-field configurations. One
obtains
\beqa
\!\!\!\!\!\!\overline{G}(t,\r_1-\r_2)&=&
\exp\left[\,\frac{i}{2}\int_{C_+} d^4x\int_{C_+}d^4y\,J_0(x) D_{00}(x-y)
J_0(y)\right]\,,\nonumber\\
&=&
\exp\!\left[\,i\,g^2\int_0^t ds \int_0^t ds'\left(D_{00}(s-s',\0)-
D_{00}(s-s',\r_1-\r_2)\right)\right],\nonumber\\ \label{eq:2QED}
\eeqa
where $ D_{\mu\nu}(x-y)$ is the real-time HTL photon propagator satisfying
the KMS condition.
The large-time behavior of the evolution equation is still given by Eq.~(\ref{eq:2nrtinfty}), with $V_{\infty}$ related  to the static  propagator, as in Eq.~(\ref{Vinftyscalar}).  The real-time photon propagator in Fourier space is given in Eq.
(\ref{eq:QEDsp}) and, 
for $\omega\!=\!0$, 
\beq\label{eq:w0QED}
D_{00}(\omega=0,\q)=\frac{-1}{\q^2+m_D^2}+i\frac{\pi m_D^2 T}
{|\q|(\q^2+m_D^2)^2}\,.
\eeq
The effective potential reads then:
\beq
V_\infty(\r_1\!-\!\r_2)\equiv g^2\int\frac{d\q}{(2\pi)^3}\left(1-e^{i\q\cdot
(\r_1-\r_2)}\right)\left[\frac{1}{\q^2+m_D^2}\!-i\frac{\pi m_D^2 T}
{|\q|(\q^2+m_D^2)^2}\right],\label{eq:impot}
\eeq
which  agrees with that found  in \cite{lai1}. Note that, as it should, the (real part of) the potential between
the quark and the anti-quark is attractive and screened.
Furthermore, contrary to what
happened  in the scalar-meson model of the previous section, the potential develops
an imaginary part. This imaginary part originates from that of the HTL photon self-energy (see App. B) and,   physically, results  from the collisions between the light fermions of the hot medium and  the heavy quarks (see App. C and below). 
Apart from a difference in the sign of the self-energy contributions,  the real part is identical  to that found in the scalar
case.  The angular integration in the imaginary part can
be easily performed and one is left with (adopting the same notation as in Ref.
\cite{lai1}):
\beq
V_\infty(\r_1\!-\!\r_2)=-\frac{g^2}{4\pi}\left[m_D+\frac{e^{-m_Dr}}{r}\right]
-i\frac{g^2T}{4\pi}\phi(m_Dr)\,,
\eeq
where $r\!\equiv\!|\r_1-\r_2|$ and
\beq
\phi(x)\equiv 2\int_0^\infty dz\frac{z}{(z^2+1)^2}
\left[1-\frac{\sin(zx)}{zx}\right]\,.
\eeq 
Thus $\overline G$ behaves as a damped exponential: 
\beq\label{eq:t_infty_QED}
\overline{G}(t,\r_1-\r_2)\!\underset{t\to+\infty}{\sim}\!
\exp\left[i\frac{g^2}{4\pi}\left(\!m_D\!+\!\frac{e^{-m_Dr}}{r}\right)\!t\right]
\!\exp\left[-\frac{g^2T}{4\pi}\phi(m_Dr)t\right].
\eeq
The damping factor increases with the temperature, as  shown in Fig. \ref{fig:gamma}: the function $\phi(x)$ vanishes
for $x\!=\!0$ and increases monotonously, approaching 1 as $x\!\to\!+\infty$. 
In the $r\!\to\!\infty$ limit the contributions
of the two heavy fermions to the pair propagator  factorize and, accordingly,
the damping factor in Eq. (\ref{eq:damping}) results from the sum of two
equal terms (for $Q$ and $\bar{Q}$ respectively):
\beq
\lim_{x\to +\infty}\phi(x)=1\,\Longrightarrow\,\gamma_Q\!=\!
\gamma_{\bar{Q}}\!=\!\frac{g^2T}{8\pi}\,,
\eeq
which coincides with the $v\!\to\!0$ limit of the heavy fermion damping factor
in Eq. (3.6) of \cite{pisar}, after dropping the Casimir factor (the
calculation of Ref.~\cite{pisar} being done in hot QCD).
Thus,  the collisional damping rate is  most important when the heavy quarks are far apart. When they get closer together, interference occurs with the process by which the exchanged photon gets absorbed by the light particles in the heat bath. When the $Q\bar Q$ separation vanishes, this interference is completely destructive and kills the imaginary part.\footnote{We therefore differ in our interpretation from that suggested in Ref.~\cite{lai5}: the ``disappearance'' of the exchange meson is in fact acting against the major contribution coming from the individual damping factors of the heavy fermions.}
\begin{figure}[!tp]
\begin{center}
\includegraphics[clip,width=0.6\textwidth]{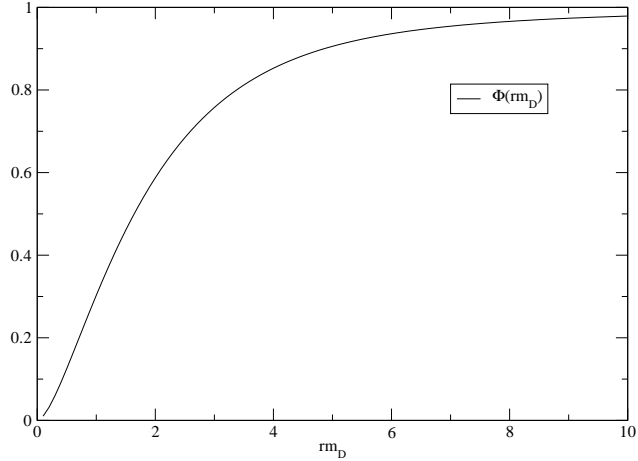}
\caption{The dependence on the $Q\overline{Q}$ separation of the function $\phi$
describing the damping of the in-medium two-particle propagator.}
\label{fig:phi} 
\end{center}
\end{figure}

In order to get a feeling for the magnitude of the effect of the imaginary part on the $Q\bar Q$ propagation in a hot medium, we have plotted  in Fig. \ref{fig:gamma} the
behavior of the damping factor
\beq
\gamma\equiv\frac{g^2T}{4\pi}\phi(m_Dr)\label{eq:damping}
\eeq
arising from Eq. (\ref{eq:t_infty_QED}), as a
function of the temperature. We shall consider  the
propagation of a heavy fermion pair in a hot QED plasma (identifying  
$g^2/4\pi\!\equiv\!\alpha_{\rm QED}$) consisting of photons,
electrons and positrons. In order to deal with some sensible numbers we take the
heavy fermion pair as being separated by a distance equal to mean value of
the radius of a $\mu^+\mu^-$ atom in its ground state, namely:
\beq
r=\langle r\rangle_{1S}=\frac{3}{2}a_{\rm Bohr}\equiv\frac{3}{2}\frac{1}
{\mu\alpha_{\rm QED}}\approx 3.89\,{\rm MeV}^{-1}\,,
\eeq
$\mu$ being the reduced mass of the $\mu^+\mu^-$ pair. One sees in Fig.~\ref{fig:gamma} that the damping is quite substantial. Actually the real growth
of the damping factor with the temperature should be even faster, since one
expects that as the temperature increases the effective potential gets more
and more screened leading to a lower binding energy and a more spread
wave-function. This would entail (treating the imaginary-part as a small
perturbation) a larger value for $\langle\gamma\rangle_{1S}$, since the
function $\phi$ in Eq. (\ref{eq:damping}) increases monotonically with the
$Q\overline{Q}$ separation.
A more refined estimate of the in-medium quarkonium decay-width has been given recently 
in \cite{lai1}.

Turning now to the Euclidean correlator, we note that the  analytic continuation to imaginary-time can be performed by following the same steps as in the scalar case. One then easily obtains 
{\setlength\arraycolsep{1pt}
\beqa
\overline{G}(-i\beta,\r_1-\r_2)
&=&\exp\left\{\beta g^2\!\int\frac{d\q}{(2\pi)^3}
\!\left(1\!-\!e^{i\q\cdot(\r_1-\r_2)}\right)\!\left(\frac{-1}{\q^2}\!+\!
\int\frac{dq^0}{2\pi}
\frac{\rho_{00}(q^0,\q)}{q^0}\right)\!\right\}\nonumber\\
{}&=&\exp\left\{-\beta g^2\!\int\frac{d\q}{(2\pi)^3}
\left(1\!-\!e^{i\q\cdot(\r_1-\r_2)}\right)\frac{1}{\q^2+m_D^2}\right\}\,,
\eeqa}
where in the last line we have employed the sum rule given in Eq.
(\ref{eq:QEDsumrule}). 
The free-energy change occurring when the $Q\overline{Q}$ pair
is placed in the  thermal bath is then:
\beq\label{free2}
\Delta F_{Q\overline{Q}}(r,T)=
-\frac{g^2m_D}{4\pi}-\frac{g^2}{4\pi}\frac{e^{-m_Dr}}{r}\,,
\eeq
$m_D$ being the Debey screening mass of the photon. Note that $\overline{G}$ can also be written as
the thermal average of two Polyakov lines (we assume the average to be made over Euclidean fields $A_4^E(\tau)$ such that $A_0(-i\tau)\!\equiv\!iA_4^E(\tau)$):
\begin{multline}\label{PLC}
 \overline{G}(-i\beta,\r_1\!-\!\r_2)=\langle\exp\!\left(\!ig\!\int_0^\beta \!d\tau'A_4^E(\tau',\r_1)\!\right)
\exp\!\left(\!-ig\int_0^\beta\!d\tau'A_4^E(\tau',\r_2)\!\right)\rangle .
\end{multline}
We recover in Eq.~(\ref{free2}) the expected lowest order perturbative result
for the $Q\overline{Q}$
free energy.
However, comparing Eq. (\ref{free2}) with Eq. (\ref{eq:t_infty_QED})
(which refers to the real-time propagation problem),
one can see that considering only the imaginary-time result at $\tau=\beta$
one misses the damping
contribution arising from the collisions with the particles of the thermal
bath.

\section{Conclusions}%
\label{sec:concl}   %

We have shown in this paper that the real-time propagator
of a (infinitely) heavy $Q\bar{Q}$ pair  crossing a hot medium obeys, at large times, a 
Schr\"odinger  equation. We have evaluated the effective potential entering this Schr\"odinger  equation using simple models for the medium in which the $Q\bar{Q}$ propagates, as well as for the interaction between the medium and the heavy fermions. Our results corroborate those of Refs.~\cite{lai1,lai2} and put them in a  broader perspective. 
In agreement with \cite{lai1,lai2,lai3}, we have found that
in the QED case the real-time potential develops an imaginary part, for which we proposed an interpretation in terms of the collisions of the heavy quarks with the light fermions of the thermal
bath. We emphasized that the damping is largest when the heavy quarks are far apart,  where it coincides with the damping  of heavy fermions that have been calculated long ago. At small separation, interference effects strongly suppress the damping in the correlator.

This effect of the collisions on the motion of the $Q\bar{Q}$, and the fate of its possible bound states, is distinct from the process of gluo-dissociation  that was first considered by Bhanot and Peskin \cite{pesk1,pesk2}, and later by \cite{oh}. The gluo-dissociation
reaction can be viewed as the LO pQCD result for the quarkonium break-up.
In \cite{wong3} the latter was the process considered in the study of 
collisional melting of charmonium in the QGP (namely the collision with a
real gluon of the thermal bath). It was also shown that the above reaction
could be considered as the analog of the deuteron photo-dissociation reaction,
whose cross section was evaluated in \cite{blatt}. The outcome of the
calculation of the gluo-dissociation process is a width of the charmonium which decreases with increasing temperature: this due to the fact that,
as the binding energy gets lower and lower, the
cross section is peaked at values of the gluon energy corresponding to
very small phase space.
The origin of the imaginary part of the potential discussed in this paper is closer to the so-called \emph{quasifree} destruction mechanism explored in 
\cite{rapp3} (the scattering off light partons of
the thermal bath) and which leads to a collisional width for the quarkonium in the QGP
that increases with the temperature \cite{rapp3,wong5}.

A second aspect of our analysis concerns the relation between the real-time and the Euclidean propagators. 
One finds that the real part of the effective potental governing the large-time
behavior of the heavy quark propagator can indeed be extracted from the
Euclidean correlator calculated for  imaginary time $\tau=\beta$, where $\beta$ is the inverse temperature.  In particular we found that the real part of the effective
potential coincides with the  $Q\overline{Q}$ free-energy.
However   we showed that the Polyakov-line correlator is
``blind'' to damping effects giving rise to the imaginary part of the effective
potential.

A limitation of the present study is of course related to the fact
that it has been performed in the $M\!\to\!\infty$ limit for the heavy
quarks. 
In \cite{lai1,lai2,lai3} the results obtained in this limit where
directly employed to address also the finite mass case,
plugging the potential into a Schr\"odinger equation that contains the kinetic energies of the heavy fermions. This was used
in particular in order
to reconstruct, starting from the popagator $G^>(t)$, the in-medium spectral
function of the quarkonium. 
It should be possible however to extend the present analysis to treat the finite mass case using the techniques of effective field theories that have been developed in the $T\!=\!0$ case~\cite{bram}.
Note finally  that in dealing with the $M\!\to\!\infty$ case one could drop contributions arising from the exchange of
magnetic photons.
While we expect magnetic photons to remain subdominant when $M$ is finite but large, their contribution may generate finite temperature IR divergences (arising from the lack of screening for the magnetostatic modes) which can be cured by the method presented in~\cite{BNbla}.

\appendix
%
\section{HTL scalar self-energy}\label{app:scalarHTL}
The  scalar self-energy is defined through  the Dyson equation
\beq
\Delta{-1}(i\omega_l,\q)=\Delta_0^{-1}(i\omega_l,\q)+\Pi(i\omega_l,\q)
\eeq
written here in imaginary time, with $\omega_l\!=\!2\pi l/\beta$ and
the free Matsubara propagator $\Delta_0^{-1}(i\omega_l,\q)=
\omega_l^2+\q^2+m^2$. 
To lowest order in $g$,  $\Pi(i\omega_l,\q)$ is given by the one-loop diagram involving  light 
fermions:
\beq
\Pi(i\omega_l,\q)=N_{\rm dof}\,g^2\frac{1}{\beta}\sum_{n=-\infty}
^{+\infty}\int
\frac{d\p}{(2\pi)^3}{\rm Tr}[S_F(i\omega_n,\p)S_F(i\omega_n-i\omega_l,\p-\q)]
\,,
\eeq
where the factor
$N_{\rm dof}$  counts the light fermion internal degrees of freedom other than spin (flavor, color).
After performing the trace over the Dirac indices, and the sum over the Matsubara frequencies,  one can perform the analytical continuation 
$i\omega_l\to q^0$, with $q^0$ off the real axis. Defining 
$\p_1\!\equiv\!\p$, $\p_2\!\equiv\!\p-\q$, $\epsilon_1\!\equiv\!p_1$ and 
$\epsilon_2\!\equiv\!p_2$, $n(\epsilon_i)=1/({\rm e}^{\beta\epsilon_i}+1)$, one finds:
\begin{multline}\label{eq:fullscalself}
\Pi(q^0,\q)=N_{\rm dof}\,g^2\int\frac{d\p_1}{(2\pi)^3}
\frac{1}{\epsilon_1\epsilon_2}\times\\
\times\left\{(\epsilon_1\epsilon_2-\p_1\cdot\p_2)\frac{n(\epsilon_2)-
n(\epsilon_1)}{q^0-(\epsilon_1-\epsilon_2)}+(\epsilon_1\epsilon_2
+\p_1\cdot\p_2)\frac{1-n(\epsilon_1)-n(\epsilon_2)}{q^0-(\epsilon_1+
\epsilon_2)}+\right.\\
-\left.(\epsilon_1\epsilon_2-\p_1\cdot\p_2)\frac{n(\epsilon_2)-n(\epsilon_1)}
{q^0+(\epsilon_1-\epsilon_2)}-(\epsilon_1\epsilon_2+\p_1\cdot\p_2)
\frac{1-n(\epsilon_1)-n(\epsilon_2)}{q^0+(\epsilon_1+\epsilon_2)}\right\}\,.
\end{multline}
The Hard Thermal 
Loop approximation \cite{lb} consists in 
the following kinematical approximation, valid when  $q_0,|\q|\ll T$:
\beq\label{eq:kinapprox}
\epsilon_2\simeq p-q\cos\theta\,,\qquad n(\epsilon_2)-n(\epsilon_1)\simeq
(\epsilon_2-\epsilon_1)\left.\frac{\partial n}{\partial \epsilon}\right|_p\!
\simeq\!-q\cos\theta\left.\frac{\partial n}{\partial \epsilon}\right|_p\,,
\eeq 
where $\theta$ is the angle between $\p$ and $\q$.
In this kinematical regime the first and the third terms within the parenthesis in 
Eq. (\ref{eq:fullscalself}) do not contribute. 
The remaining terms  contain contributions that are UV divergent term but independent of the  temperature or of the momentum, and can be absorbed into the renormalization of the meson mass. 
One finally gets~\cite{tho} 
\beq
\Pi(q^0,\q)=4N_{\rm dof}\,g^2\,\int\frac{d\p}{(2\pi)^3}\frac{n(p)}{p}=
N_{\rm dof}\frac{g^2T^2}{6}\,.
\eeq
%
\section{The HTL real-time photon propagator}\label{app:QED}%

The one-loop longitudinal photon self-energy reads (for simplicity we
assume that the coupling with the electromagnetic field is the same for all
the species of light fermions):
\beq
\Pi_{00}(i\omega_l,\q)=N_{\rm dof}\,g^2\frac{1}{\beta}\sum_{n=-\infty}
^{+\infty}\int
\frac{d\p}{(2\pi)^3}{\rm Tr}[\gamma^0 S_F(i\omega_n,\p)\gamma^0
S_F(i\omega_n-i\omega_l,\p-\q)]
\,,
\eeq
leading to, after summation of the Matsubara frequencies, and analytic continuation
\begin{multline}\label{eq:QEDself}
\Pi_{00}(q^0,\q)=N_{\rm dof}\,g^2\int\frac{d\p_1}{(2\pi)^3}
\frac{1}{\epsilon_1\epsilon_2}\times\\
\times\left\{(\epsilon_1\epsilon_2+\p_1\cdot\p_2)\frac{n(\epsilon_2)-
n(\epsilon_1)}{q^0-(\epsilon_1-\epsilon_2)}+(\epsilon_1\epsilon_2
-\p_1\cdot\p_2)\frac{1-n(\epsilon_1)-n(\epsilon_2)}{q^0-(\epsilon_1+
\epsilon_2)}+\right.\\
-\left.(\epsilon_1\epsilon_2+\p_1\cdot\p_2)\frac{n(\epsilon_2)-n(\epsilon_1)}
{q^0+(\epsilon_1-\epsilon_2)}-(\epsilon_1\epsilon_2-\p_1\cdot\p_2)
\frac{1-n(\epsilon_1)-n(\epsilon_2)}{q^0+(\epsilon_1+\epsilon_2)}\right\}\,,
\end{multline}
In the HTL approximation, the dominant contributions come from the "Landau damping" process (the first and third terms in the equation above; note the difference with the scalar case), and we have
\beqa
\Pi_{00}(q^0,\q)&=&4 N_{\rm dof}\frac{g^2}{(2\pi)^2}\int_0^\infty p^2dp
\left(-\frac{\partial n}{\partial\epsilon}\right)_p\,\int_{-1}^{1}d\cos\theta
\frac{q\cos\theta}{q^o-q\cos\theta}\nonumber\\
&=&-m_D^2\left(1-\frac{x}{2}\ln\frac{x+1}{x-1}\right)\,,
\eeqa
where $x\!\equiv\!q^0/q$ and $m_D$ is the Debye screening mass
\beq
m_D^2\equiv N_{\rm dof}\frac{g^2T^2}{3}.
\eeq
 Form this one deduces in particular the imaginary part ($q_0=\omega+i\eta$)
\beq
{\rm Im} \Pi_{00}(\omega+i\eta,\q)=-m_D^2\frac{\pi\omega}{2q}\theta(q-|\omega|)
\eeq
and, after solving the Dyson equation with the non-interacting longitudinal
propagator $\left.\Delta_0^{-1}\right|_{00}\!=\!-1/\q^2$, the analytic
(off the real axis) propagator
\beq
\Delta_{00}(q^0,\q)=\frac{-1}
{\q^2+m_D\left(1-\frac{x}{2}\ln\frac{x+1}{x-1}\right)}\,.
\eeq
From the latter one gets
the longitudinal photon spectral function which allows to express the real-time
propagator as follows:
\beq\label{eq:QEDsp}
D_{00}(\omega,\q)=\frac{-1}{\q^2}+\int_{-\infty}^{+\infty}\frac{dq^0}{2\pi}
\frac{\rho_{00}(q^0,\q)}{q^0-(\omega+i\eta)}+i\rho_{00}(\omega,\q)N(\omega)\,.
\eeq

The $\omega =0$ limit of the above real time correlator can be obtained from
the sum rule \cite{lb}
\beq\label{eq:QEDsumrule}
\int_{-\infty}^{+\infty}\frac{dq^0}{2\pi}\frac{\rho_{00}
(q^0,\q)}
{q^0}=\frac{1}{\q^2}-\frac{1}{\q^2+m_D^2}\,,
\eeq
and from the low-energy behavior:
\beq\label{eq:lo_en}
\rho_{00}(\omega,\q)N(\omega)\underset{\omega\to 0}{\sim}
\frac{\pi m_D^2 \omega}{|\q|(\q^2+m_D^2)^2}
\frac{T}{\omega}\,.
\eeq
\section{Physical interpretation of the damping}\label{app:damping}

Here we  give a physical
interpretation of the processes responsible for the damping of the $Q\overline{Q}$
correlator discussed in Sec. \ref{sec:QEDtoy} .

\begin{figure}[!tp]
\begin{center}
\includegraphics[clip,width=0.6\textwidth]{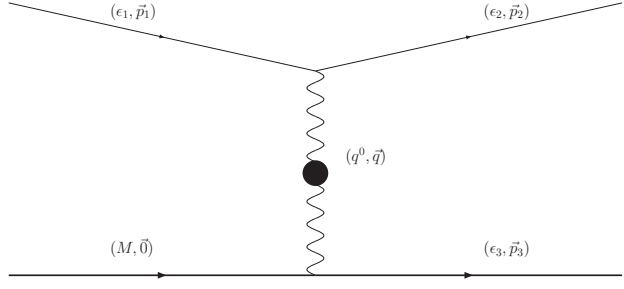}
\caption{The collision process responsible for the imaginary part of the $Q\bar Q$ correlator }\label{fig:rate} 
\end{center}
\end{figure}

We consider a heavy fermion  with four-momentum
$P\!=\!(M,\0)$  ($M\!\gg\!T$) in a hot   QED
plasma. The interaction rate with the  light fermions 
of the thermal bath is given by:
\begin{multline}\label{Gamma_rate}
\Gamma(M)=2\,\frac{1}{2M}\int_{p_1}\int_{p_2}\int_{p_3}(2\pi)^4\delta^{(4)}
(P+P_1-P_2-P_3)\times\\
\times\left[n_1(1-n_2)(1-n_3)+(1-n_1)n_2 n_3\right]\overline{|{\cal M}|^2}\,,
\end{multline}
where $\int_{p_i}\!\equiv\!\int d\p_i/(2\pi)^3 2\epsilon_i$, $n_i$ is a fermion  occupation factor and the overall
factor 2 accounts for scattering over particles and antiparticles.
The corresponding process is depicted in Fig.~\ref{fig:rate}: the
particle $1$ hits the heavy fermion and comes out with four-momentum $P_2$.
The index $3$ refers to the four-momentum of the heavy fermion after the
scattering. The second term in the parenthesis accounts for the inverse
process.When the mass of the heavy fermion is infinite,
$n_3\!=\!0$ and the inverse process does not contribute.

The modulus square of the amplitude, averaged over the spin of the incoming 
heavy fermion and summed over the spins of the other particles is easily
evaluted \cite{bi}:
\begin{multline}
\overline{|{\cal M}|^2}=8g^4 \Delta_{\mu\nu}(q)\Delta_{\rho\lambda}^*(q)
\left[P^\mu P_3^\rho+
 P_3^\mu P^\rho-g^{\mu\rho}(P\!\cdot\!P_3)+g^{\mu\rho}M^2\right]\times\\
\times\left[P_1^\nu P_2^\lambda+
 P_2^\nu P_1^\lambda-g^{\nu\lambda}(P_1\!\cdot\!P_2)\right]\,.
\end{multline}
In the above $\Delta_{\mu\nu}$ is the HTL resummed photon propagator, which
we take in the Coulomb gauge. Furthermore, in the large $M$ limit, 
 we can keep only the
contribution   arising from the exchange of a longitudinal
photon and ignore that from the transverse photons. Finally, because the interaction is dominated by the exchange of
soft momenta, 
$\epsilon_3\!\approx\!M$ and $\p_1\!\approx\!\p_2$. Under the above
approximations one can write:
\beq
\overline{|{\cal M}|^2}=8g^4|\Delta_{00}(q)|^2(2M^2)(2\epsilon_1\epsilon_2)\,.
\eeq
The statistical factor can be rewritten as \cite{bi}:
\beq
n_1(1-n_2)(1-n_3)+(1-n_1)n_2 n_3=(n_1-n_2)(1+N(q_0)-n_3)\,.
\eeq
In the present kinematical conditions, where 
$p_1\!\sim\!p_2\!\sim\!T$, and  $q^0=\epsilon_2-\epsilon_1\ll T$, one can write:
\beq
(1+N(q_0)-n_3)\simeq\frac{T}{q^0}
\eeq
and
\beq
(n_1-n_2)(1+N(q_0)-n_3)\simeq -\frac{dn}{dp_1}(\epsilon_2-\epsilon_1)\frac
{T}{q^0}\simeq -T\frac{dn}{dp_1}\,.
\eeq
It is then convenient to change the integration variables as follows \cite{bi}:
\begin{multline}\label{eq:change}
\int\frac{d\p_2}{(2\pi)^3}\int\frac{d\p_3}{(2\pi)^3}(2\pi)^4\delta^{(4)}
(P+P_1-P_2-P_3)=\\
\left.=\int\frac{d\q}{(2\pi)^3}\int\frac{dq^0}{2\pi}2\pi\delta(q^0)\,2\pi\delta
(q^0-\v_1\cdot\q)=\int\frac{d\q}{(2\pi)^3}\delta
(\v_1\cdot\q)\right|_{q^0=0}\,,
\end{multline}
where the two Dirac-distributions account for energy conservation at the
interaction vertices. The first one expresses the fact that the energy transfer to the heavy fermion $\sim \q^2/2M$ is negigible.  The second  expresses
the conservation of energy at the vertex with the light fermions, in the HTL approximation where $
q^0=\epsilon_2-\epsilon_1\approx \v_!\cdot\q$. One can use  Eq. (\ref{eq:change}) to perform 
the angular integration
over $\p_1$ in Eq.~(\ref{Gamma_rate}). One is left with:
\beqa
\Gamma&=&g^2T\frac{2g^2}{\pi}\int_0^\infty p_1^2 dp_1
\left(-\frac{dn}{dp_1}\right)\int\frac{d\q}{(2\pi)^3}\frac{1}{q}
|\Delta_{00}(0,\q)|^2\,,
\nonumber\\ &=& g^2T\int\frac{d\q}{(2\pi)^3}\frac{\pi m_D^2}{(\q^2+m_D^2)^2 q}\,,
\eeqa
which is precisely the contribution to the $Q\overline{Q}$ damping factor
arising from processes involving a single heavy fermion. 

\end{document}